\documentclass{aa}

\usepackage{graphicx}
\usepackage{txfonts}
\usepackage{multirow}
\usepackage{amssymb}
\usepackage{units}
\usepackage{float}
\usepackage{natbib}
\bibpunct{(}{)}{;}{a}{}{,}

\usepackage[colorlinks=true, citecolor=blue]{hyperref}
\usepackage[dvipsnames]{xcolor}

\makeatletter
\renewcommand*\aa@pageof{, page \thepage{} of \pageref*{LastPage}}
\makeatother

\begin{document}

   \title{Interstellar dust along the line of sight of GX 3+1}
 
   \titlerunning{Mg \& Si K-edges of GX 3+1}
   \authorrunning{D. Rogantini, E. Costantini et al.}

   \subtitle{}

   \author{D. Rogantini,
          \inst{1,2}
          E. Costantini,\inst{1,2}
          S.T. Zeegers,\inst{3}
          C.P. de Vries,\inst{1}
          M. Mehdipour,\inst{1}
          F. de Groot,\inst{4}
          H. Mutschke,\inst{5}
          I. Psaradaki,\inst{1}
           \and
          L.B.F.M. Waters,\inst{1,2}
          \
          }

   \institute{SRON Netherlands Institute for Space Research, Sorbonnelaan 2, 3584 CA Utrecht, the Netherlands\\
              \email{d.rogantini@sron.nl}
             \and 
             Anton Pannekoek Astronomical Institute, University of Amsterdam, P.O. Box 94249, 1090 GE Amsterdam, the Netherlands
             \and
             Academia Sinica Institute of Astronomy and Astrophysics, 11F of AS/NTU  Astronomy-Mathematics Building, No.1, Section 4, Roosevelt Rd, Taipei10617, Taiwan, ROC
             \and
             Debye Institute for Nanomaterials Science, Utrecht University, Universiteitweg 99, 3584 CG Utrecht, the Netherlands
             \and
             Astrophysikalisches Institut und Universit$\ddot{\text{a}}$ts-Sternwarte (AIU), Schillerg{\"a}{\ss}chen 2-3, 07745 Jena, Germany
                          }
    \date{Date: \emph{-}}

 
  \abstract
   {
   Studying absorption and scattering of X-ray radiation by interstellar dust grains allows us to access the physical and chemical properties of cosmic grains even in the densest regions of the Galaxy.
   }
   {
   We aim at characterising the dust silicate population which presents clear absorption features in the energy band covered by the \textit{Chandra} X-ray Observatory. Through these absorption features, in principle, it is possible to infer the size distribution, composition, and structure of silicate in the interstellar medium. In particular, in this work we investigate the magnesium and silicon K-edges.
   }
   {
   By using newly acquired synchrotron measurements, we build X-ray extinction models for fifteen dust candidates. These models, adapted for astrophysical analysis, and implemented in the \textsc{Spex} spectral fitting program, are used to reproduce the dust absorption features observed in the spectrum of the bright low mass X-ray binary GX 3+1 which is used as a background source.
   }
   {
   With the simultaneous analysis of the two edges we test two different size distributions of dust: one corresponding to the standard Mathis-Rumpl-Nordsieck model and one considering larger grains ($n(a) \propto a_i^{-3.5}$ with $0.005<a_1<0.25$ and $0.05<a_2<0.5$, respectively, with $a$ the grain size). These distributions may be representative of the complex Galactic region towards this source. We find that up to $70\%$ of dust is constituted by amorphous olivine. We discuss the crystallinity of the cosmic dust found along this line of sight. Both magnesium and silicon are highly depleted into dust ($\delta_{Z} = 0.89\ \rm{and}\ 0.94$, respectively) while their total abundance does not depart from solar values.
   }
   {
   }


   \maketitle
%


\section{Introduction}
Magnesium is an essential element for life mainly because of its wide presence in the basic nucleic acid chemistry of all cells of all known living organisms \citep{Cowan95}. This is not surprising, given the relative large fraction of Mg in the interstellar medium (ISM). The solar photospheric abundance of magnesium is $\log A_{\rm{Mg}} = 7.54 \pm 0.06$\footnote{The abundances are given in logarithmic scale relative to a hydrogen column density $N_{\rm{H}} = 10^{12}$. Explicitly for magnesium we have
$$\log A_{\rm{Mg}} = 12 + \log (N_{\rm Mg}/N_{\rm H})\ ,$$ where $N_{\rm{Mg}}$ is the indicate the magnesium column density.} \citep[the ninth element in order of mass abundance;][]{Lodders10} and it is consistent with the chondrite composition in the solar nebula \citep{Anders89}. Magnesium is primarily synthesised in Type Ia supernovae and in core-collapse supernovae \citep{Heger10}, and it is present in quiescent stellar outflows during the asymptotic giant branch (AGB) phase of their evolution \citep{vandenHoek97}.\\
In the interstellar medium magnesium is significantly depleted into the solid phase. The depletion is parametrised by the depletion index which refers to the underabundance of the gas-phase element with respect to its standard abundance, resulting from its inclusion in cosmic grains. This term depends on the environment properties showing three typical patterns as a function of density, turbulence, and Galactic latitude \citep{Jones00,Whittet02}: high element depletions are found in dense, quiescent regions in the Galactic plane.\\
The mean value of depletion index for magnesium in the diffuse clouds is $D_{Mg} = -1.10$ with fractional depletion\footnote{The fractional depletion, often expressed as a percentage, is defined as $\delta_{\rm{X}} = 1- 10^{D_{\rm{X}}}$ where the depletion index $D_{\rm{X}}$ is evaluated comparing the abundance of the gas-phase element X with respect to its standard solar reference abundance:
$
D_{\rm{X}} = \log \big\{ \frac{N_{\rm{X}}}{N_{\rm{H}}} \big\} - log \big\{ \frac{N_{\rm{X}}}{N_{\rm{H}}} \big\}_{\odot}.
$
}
in the range $  \delta_{\rm{Mg}} = 0.85-0.95$ \citep[see e.g.,][]{Savage96,Jenkins09}.
Together with silicon, magnesium is almost completely included in silicate grains. Silicate dust is of great interest to astronomers due to its prevalence in many different astrophysical environments, including the diffuse interstellar medium, protoplanetary disks around young stars, evolved and/or massive stars \cite[e.g., AGB stars, red supergiant stars, supergiant Be stars, see][and references therein]{Henning10}, and even in the immediate environments of active galactic nuclei \cite[i.e.][]{Markwick-Kemper07,Mehdipour18}. \\
The physical and chemical properties of silicates in the interstellar medium have traditionally been studied through infrared spectroscopy. The broad and smooth infrared features at 10 and 18 $\mu$m are attributed to Si-O stretching and O-Si-O bending modes of cosmic silicates in amorphous state \citep{Henning10}. However, it is still not known exactly what composition or structure (i.e. dust size, crystallinity) characterise these dust grains, or how these properties change as a function of the galactic environment \citep{Speck15}. The shape, position and the width of the two bands depend on multiple factors often difficult to disentangle, such as the level of $\rm{SiO}_{4}$ polymerization \citep{Jager03b}, the Fe content \citep{Ossenkopf92}, crystallinity \citep{Fabian00}, particle size \citep{Li01b}, and particle shape and size of the interstellar dust grains \citep{Voshchinnikov08,Mutschke09}.\\

\noindent
X-ray observations provide a powerful and direct probe of cosmic silicates and interstellar dust in general \citep{Draine03}. The cosmic grains interact with the X-ray radiation by absorbing and scattering the light. In particular the X-ray energy band contains the absorption edges of the most abundant metals. Several works \citep{Lee09, Costantini12, Pinto13,Valencic13,Corrales15,Zeegers17,Bilalbegovic18,Rogantini18} have already shown how these absorption edges allow us to study in detail the chemical and physical properties of the dust grains. Differently from the gas phase, the interaction between X-rays and solid matter modulates the post-edge region and imprints characteristic features. These features, named X-ray Absorption Fine Structures \citep[XAFS, see][for a detail theoretical explanation]{Bunker10} are characteristic of the chemical species present in the absorber. They are unique fingerprints of dust. Moreover, these features are sensitive to the crystalline order of the grains. The peak on the pre-edge is due to the scattering interference between the X-rays and the grains. \cite{Zeegers17} and \cite{Rogantini18} have shown how this scattering peak is sensitive to the grain size and how it allows us to investigate the dust geometry in different environments for the Si and Fe absorption edges, respectively. XAFS is often divided into two regimes: the X-ray absorption near-edge structure (XANES) which extends from about 5 to 10 eV below the K- or L- edge threshold energy to about 30 eV above the edges; and the extended X-ray absorption fine structure (EXAFS) which extends from $\sim 5$ eV above the K- or L- edge energy to some hundreds eV \citep{Newville04}.\\

\noindent
In order to determine the nature of dust grains in space we first accomplish laboratory measurements of dust analogue minerals whose chemical compositions are well characterised to derive optical functions of minerals predicted to occur in space. Afterwards, we match the positions, widths, and strengths of observed spectral absorption features with those seen in the laboratory spectra.\\
We use low mass X-ray binaries (LMXBs) as background sources to illuminate the interstellar dust along the line of sight. The standard spectrum of this X-ray source class does not usually present emission lines, which may confuse the absorption spectrum, and it is characterised by a high continuum flux. As they are distributed along the Galactic plane, the X-ray emission of LMXBs allows us to investigate a large range of column densities including those crossing dense interstellar dust environments of the Galaxy.\\

\noindent
In this paper we characterise simultaneously the extinction by Mg- and/or Si- bearing cosmic grains along the line of sight of a bright LMXB. We use multiple-edge extinction models that we build from synchrotron measurements. Here, we focus on the Mg K-edge. In Section \ref{sec:data} we present the relative extinction cross sections of a set of physically motivated compounds. The Si K-edge profiles are taken from the works of \cite{Zeegers17,Zeegers19}. In Section \ref{sec:esa} we present the bright LMXB, GX 3+1. For the analysis of its spectrum, we use \textsc{Spex} version 3.04.00 \citep{Kaastra96,Kaastra17b}. The source presents a line-of-sight hydrogen column density \citep[$N_{\rm{H}}\sim1.6\times 10^{22}\,\rm{cm^{-2}}$][]{Oosterbroek01} and a flux \citep[$F_{2-10\,\rm{keV}} \sim 4 \times 10^{-9}\ \rm{erg}\ \rm{s}^{-1}\ \rm{cm}^{-2}$,][]{Oosterbroek01} ideal to study the cold absorbing medium through the two extinction edges of interest. Although the spectrum of GX 3+1 is well known in the literature, the absorption by cold interstellar dust has never been studied in detail for this source. The results of the Mg and Si edges analysis are discussed and summarised in Section \ref{sec:discussion} and \ref{sec:conclusion}, respectively.


\section{Laboratory data analysis}{}{}
\label{sec:data}

\subsection{The sample}
\label{sec:sample}

The laboratory sample-set belongs to a larger synchrotron measurement campaign already presented by \cite{Costantini13} and \cite{Zeegers19}. In the first part of Table \ref{tab:sample} we present all the laboratory samples for which we measured the Mg K-edge. We report their chemical formula, form, and origin. Some of them are already presented in previous works \citep{Zeegers17, Rogantini18}. In this work we refer to the models used in the spectroscopic analysis of the astronomical source using the \#Mod indexes of Table \ref{tab:sample}.\\
In order to reproduce laboratory analogues of astronomical silicates, we have selected our samples taking into account both the two main stoichiometric classes, olivine and pyroxene. Both classes share the same building block represented by the silicate tetrahedron, SiO$_4$. This is a four-sided pyramid shape with an atom of oxygen at each corner and silicon in the middle. However, the spatial disposition of the tetrahedron is different \citep{Panchuk17}: olivine shows a structure composed of isolated tetrahedra whereas pyroxene is an example of a single-chain silicate where adjacent tetrahedron share one oxygen atom.\\

In our sample set we consider pyroxenes and olivine with varying Mg-to-Fe ratio. Olivine can be pure $\rm{Mg}_{2}\rm{Si}\rm{O}_{4}$ (forsterite) or $\rm{Fe}_{2}\rm{Si}\rm{O}_{4}$ (fayalite) or some combination of the two, written as $(\rm{Mg},\rm{Fe})_{2}\rm{Si}\rm{O}_{4}$. Pyroxene can be Mg-pure $\rm{Mg}\rm{Si}\rm{O}_{3}$ (enstatite) or Fe-pure $\rm{Fe}\rm{Si}\rm{O}_{3}$ (ferrosilite) or combination of the two.  The nomenclature En(x)Fs(1-x) indicates the fraction of iron (or magnesium) included in the compound. "En" stands for enstatite and "Fs" for ferrosilite. These silicate compounds are present in both crystalline and amorphous forms. We complete our sample set adding spinel, a Mg-bearing compound which crystallise in the cubic crystal system formed by oxygens whereas Mg and Al atoms sit in tetrahedral and octahedral sites in the lattice \citep{Mutschke98}. Spinel has been observed in chondritic meteorite with pre-solar composition and it has been produced by gas outflows of red giant stars \citep{Zinner05}. 
%
%
\begin{table}
\caption{List of samples in our set with their relative chemical formula, form, origin and reference index. In the top part of the table we list the compounds for which the Mg K-edge was analysed in the present work.
        }            
\label{tab:sample}      
\centering
\renewcommand{\arraystretch}{1.1}
\begin{tabular}{r l l c}        
\hline\hline          
 Name & Chemical formula & Form & \#Mod \\   
\hline                       
\noalign{\vskip 1.0mm}
  Enstatite\tablefootmark{a}    		& MgSiO$_3$  								              & amorphous		& 1  \\
  Enstatite\tablefootmark{b}	      & MgSiO$_3$        							          & crystalline	& 2  \\
  Forsterite\tablefootmark{c}   		& Mg$_{2}$SiO$_4$         	 				      & crystalline	& 3  \\
  Hypersthene\tablefootmark{d}      & Mg$_{1.502}$Fe$_{0.498}$Si$_2$O$_6$  		& crystalline	& -  \\
  Olivine\tablefootmark{a} 			    & MgFeSiO$_4$      	 						          & amorphous  	& 4  \\
  Olivine\tablefootmark{e}   		    & Mg$_{1.56}$Fe$_{0.4}$Si$_{0.91}$O$_4$ 	& crystalline	& 5  \\
  En60Fs40\tablefootmark{a}   		  & Mg$_{0.6}$Fe$_{0.4}$SiO$_3$         		& amorphous	  & 6  \\
  En60Fs40\tablefootmark{a}         & Mg$_{0.6}$Fe$_{0.4}$SiO$_3$             & crystalline & 7  \\
  En75Fs25\tablefootmark{a}   		  & Mg$_{0.75}$Fe$_{0.25}$SiO$_3$  	  		  & amorphous		& 8  \\
  En90Fs10\tablefootmark{a} 			  & Mg$_{0.9}$Fe$_{0.1}$SiO$_3$            	& amorphous		& 9  \\
  En90Fs10\tablefootmark{a}   		  & Mg$_{0.9}$Fe$_{0.1}$SiO$_3$         		& crystalline	& 10 \\
  Spinel\tablefootmark{f}  	        & MgAl$_2$O$_4$                         	& crystalline	& -  \\
  \noalign{\vskip 1.0mm}
\hline
\noalign{\vskip 1.0mm}
  Quartz\tablefootmark{g}           & SiO$_2$                                 & crystalline & 11 \\
  Quartz\tablefootmark{g}           & SiO$_2$                                 & amorphous   & 12 \\
  Quartz\tablefootmark{g}           & SiO$_2$                                 & amorphous   & 13 \\
  Fayalite \tablefootmark{g}        & Fe$_{2}$SiO$_4$                         & crystalline & 14 \\
  Magnesia \tablefootmark{h}        & MgO                                     & crystalline & 15 \\
\noalign{\vskip 1.0mm}

\hline                                   
\end{tabular} 
\tablefoottext{\scriptsize{a}}{\footnotesize{Synthesised in laboratories at AIU Jena and Osaka University;}}
\tablefoottext{\scriptsize{b}}{\footnotesize{Origin: Kiloza, Tanzania;}}
\tablefoottext{\scriptsize{c}}{\footnotesize{Commercial product (Alfa Aesar);}}
\tablefoottext{\scriptsize{d}}{\footnotesize{Origin: Paul Island, Labrador;}}
\tablefoottext{\scriptsize{e}}{\footnotesize{Origin: Sri Lanka;}}
\tablefoottext{\scriptsize{f}}{\footnotesize{Commercial product (Aldrich);}}
\tablefoottext{\scriptsize{g}}{\footnotesize{We refer to \cite{Zeegers19}, the quartz have two different stages of amorphisation}}
\tablefoottext{\scriptsize{h}}{\footnotesize{Also known as magnesium oxide, data taken from \cite{Fukushi17}.}}
\end{table}

\subsection{Synchrotron measurements}

Similarly to the Si K-edge already presented by \cite{Zeegers17,Zeegers19}, for the Mg K-edge we made use of the laboratory data that we obtained at the beamline "LUCIA" \citep[Line for Ultimate Characterization by Imaging and Absorption,][]{Flank06} at the SOLEIL facility in Paris. LUCIA is an X-ray microprobe capable of performing spatially-resolved chemical speciation via X-ray absorption spectroscopy (XAS). The $0.8-8\ \rm keV $ X-ray domain of the tunable beam gives access to the K edges of low Z elements (from sodium up to iron) with a resolving power of about 4000. The measurements of the magnesium absorption edge are part of a larger synchrotron campaign in which the absorption edges due to Al and Si were also measured \citep{Costantini19,Zeegers19}.\\
The spectrum around the Mg K-edge was taken in the fluorescence geometry detecting the "secondary" (fluorescent) X-ray emission from the sample that has been excited by bombarding it with the synchrotron radiation. X-rays are energetic enough to expel tightly-held electrons (photo-electron) from the inner orbitals (K-shell) making the electronic structure of the atom unstable. Consequently, one electron falls from a higher orbital level to the lower orbital to fill the hole left behind by a photo-electron. As a consequence, it releases fluorescent energy. This fluorescent signal can be used to derive the amount of absorption beyond the edge.\\
For each compound we took $3-4$ measurements to average the signal and smooth out the possible small instrumental oscillations. Finally, we shifted our measurements by 2.54 eV to lower energies since the undulator radiation of the synchrotron introduced a systematic shift in the monochromator. In Appendix \ref{app:mgk_shift} we describe how we determined the exact value of this energy shift.

\subsection{Extinction cross sections}
\label{sec:ext}
\begin{figure}[t]
   \centering
   \includegraphics[width=.981\hsize]{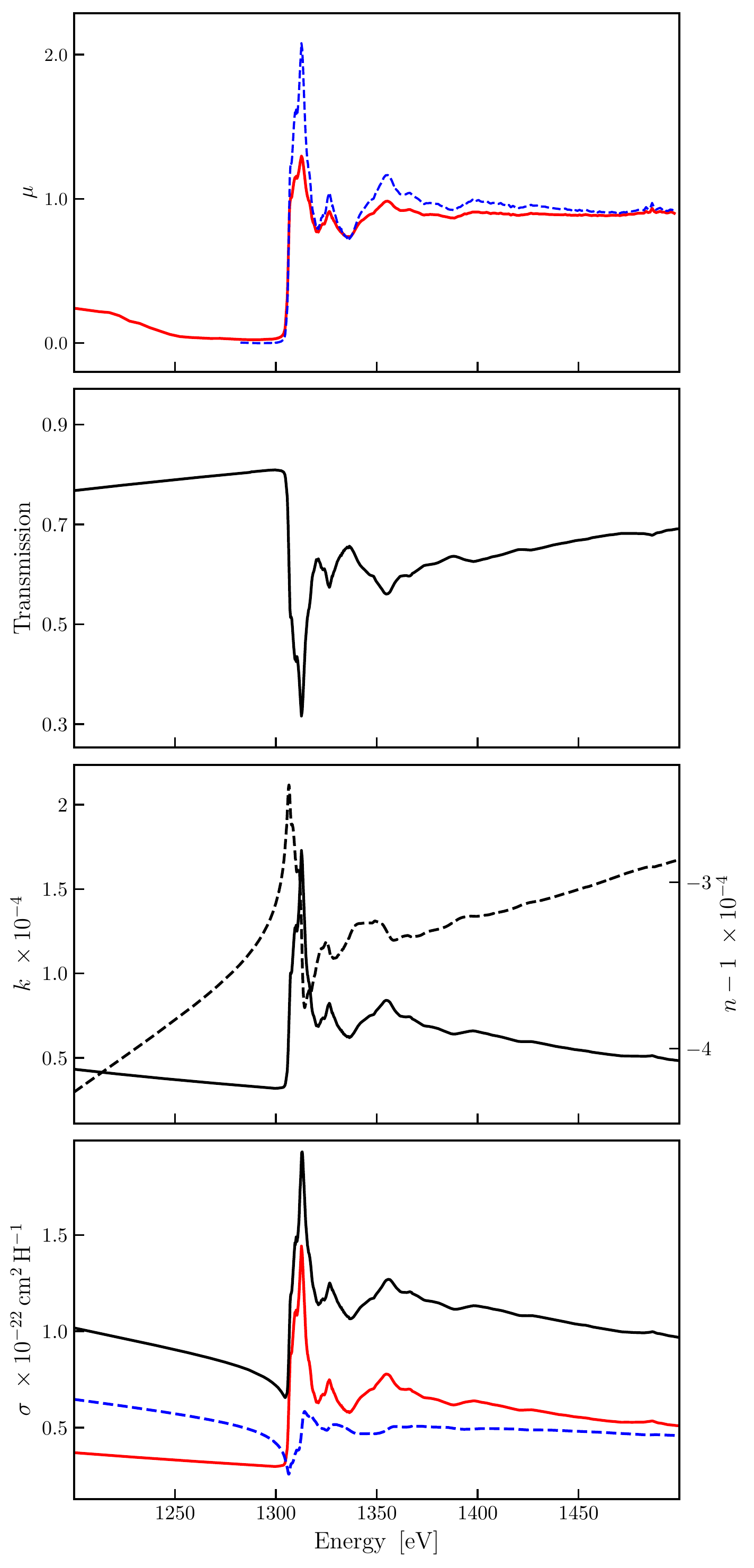}
      \caption{Representation of the data analysis for the forsterite $\rm{Mg}_2\rm{SiO}_4$. From top to bottom: $(a)\,-$ Self absorption correction: in red solid line the synchrotron raw data and in blue dashed line the signal corrected with the FLUO tool. $(b)\,-$ Transmission for a thin layer ($\tau = 0.5\, \mu\rm{m}$): the measured edge with XAFS (in black) are normalised using the tabulates values from \cite{Henke93}. $(c)\,-$ Optical constants: $k$ is represented with the solid line while $n-1$ with the dashed line (in units of $10^{-4}$). $(d)\,-$ The extinction (black solid), absorption (red solid) and scattering (blue dashed) cross sections per hydrogen nucleus for the \cite*{Mathis77} dust model.
              }
         \label{fig:data_analysis}
   \end{figure}

In order to study the attenuation of X-rays by the interstellar dust it is necessary to calculate the extinction cross section of each sample. We follow the same method already presented in \cite{Zeegers17} and \cite{Rogantini18}. Here, we summarise the procedure highlighting the most relevant steps. The main results are shown in the multiple panels of Figure \ref{fig:data_analysis}. 

   {}

\begin{description}
\item[\it{Pile-up and self-absorption correction}] - Ideally, the samples should be either sufficiently thin or sufficiently diluted for the data to be unaffected by self-absorption effect. Practically, this may not be possible and the consequence would be an incorrect peak size in the XAFS. This is due to the variation in penetration depth into the sample as the energy is scanned through the edge and the fine structure \citep{Troger92}. \\
Therefore, we correct the spectrum using the standard FLUO algorithm\footnote{\url{https://www3.aps.anl.gov/~haskel/fluo.html}}, which is part of the UWXAFS analysis package \citep{Stern95}. For comparison, we also used for the self-absorption correction the tool ATHENA\footnote{Here, ATHENA is an interactive graphical utility for XAFS data inside a the comprehensive data analysis system DEMETER (see \url{https://bruceravel.github.io/demeter/documents/Athena/index.html}). Not to be confused with the future X-ray observatory \textit{Athena}.}, obtaining the same corrected signal \citep{Ravel05}. Finally, we also correct the beamline data for any pile-up effect. For the Mg K-edge, this effect slightly distorts the region extending beyond the edge. In Figure \ref{fig:data_analysis}$a$ we compare the raw data (solid red line) and the corrected one (dashed blue line). It is necessary to ignore part of the pre-edge since the beamline was not yet stable during the measurement in this energy range.
\item[\it{Transmission}] - With the goal of determining the attenuation coefficient ($\mu$) in $\mu \rm{m}^{-1}$ necessary to calculate the refractive index of the material, we transform the absorption in arbitrary units obtained from the fluorescent measurements into transmittance. We use tabulated values of the X-ray transmission of solids provided by the Centre for X-ray Optics at Lawrence
Berkeley National laboratory\footnote{\url{http://www.cxro.lbl.gov/}} (CXRO). In order to simulate the optically thin interstellar medium condition, we choose a thickness of $0.5\, \mu \rm{m}$ (a value far below the attenuation length of each sample). Knowing the transmittance (shown in Figure \ref{fig:data_analysis}$b$) it is possible to acquire the optical constants.
\item[\it{Optical constants}] - In order to obtain the extinction cross section of a specific material, it is fundamental to derive the refractive index. It is a complex and dimensionless quantity generally defined as $m = n +ik$. The imaginary absorptive part $k$ is derived directly from the laboratory data, in specific from the transmittance signal. The real dispersive part $n$, on the other hand, can be calculated using the Kramers-Kronig relation \citep{Kronig26,Kramers27}. For this calculation we use the algorithm introduced by \cite{Watts14}. The final results are shown in Figure \ref{fig:data_analysis}$c$. For further details on the calculation of the optical constants we refer to the dedicated paragraph in \cite{Rogantini18}. 
\item[\it{Cross sections}] - To obtain the cross section from the optical constants, $n$ and $k$, we employ the anomalous diffraction theory \citep[ADT,][]{Hulst57}. This method allows to compute the absorption and the scattering by dust grains of arbitrary geometry. In this step it is important to define the grain size range of interest. We calculate the scattering, absorption and extinction cross sections (shown in Figure \ref{fig:data_analysis}$d$) for each compound using the standard MRN grain size distribution \citep[][]{Mathis77}. See also Section \ref{sec:large}.
\item[\it{Model in \textsc{Spex}}] - Finally, we implement the extinction profiles into the \textsc{Spex} fitting code adding them to the library of \texttt{amol} \citep{Pinto10}. Currently this model uses the Verner absorption curves \citep{Verner96}. We adjust the slopes of the pre- and post- edge of our extinction profiles avoiding any discontinuities between the XAFS data and the predefined curve in \textsc{Spex} \cite{Zeegers17}.
\end{description}

   \begin{figure}
   \centering
   \includegraphics[width=\hsize]{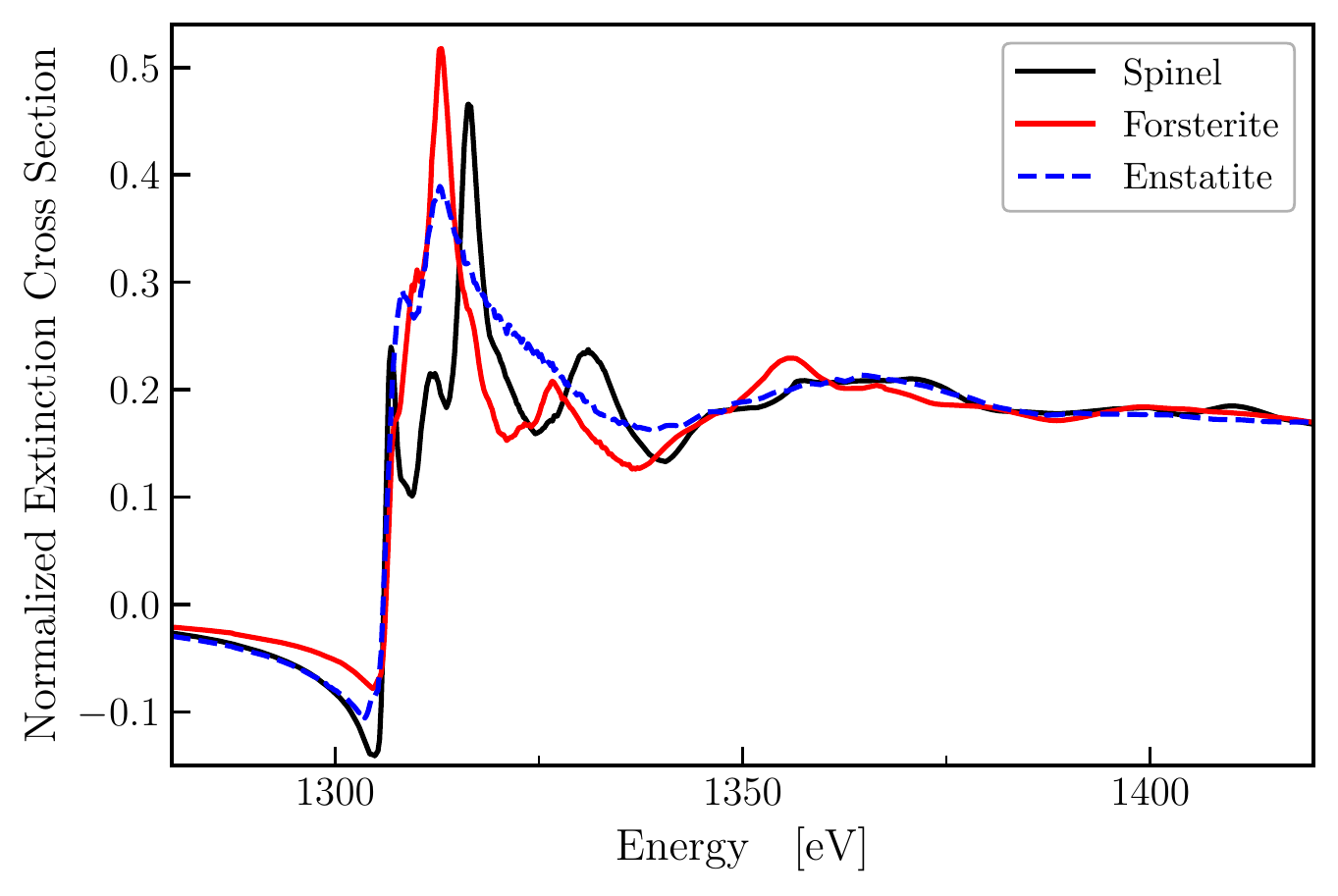}
      \caption{ The Mg K-edge model implemented in SPEX for three different chemical compounds: crystalline spinel ($\rm{MgAl}_2\rm{O}_4$), crystalline forsterite ($\rm{Mg}_2\rm{SiO}_4$), and amorphous enstatite ($\rm{Mg}\rm{SiO}_3$). The major peak of spinel in the post edge is shifted at higher energy with respect to the silicates. This is due to a different configuration of the atoms in the single unit cell. 
              }
         \label{fig:chemistry}
   \end{figure}

   \begin{figure}
   \centering
   \includegraphics[width=\hsize]{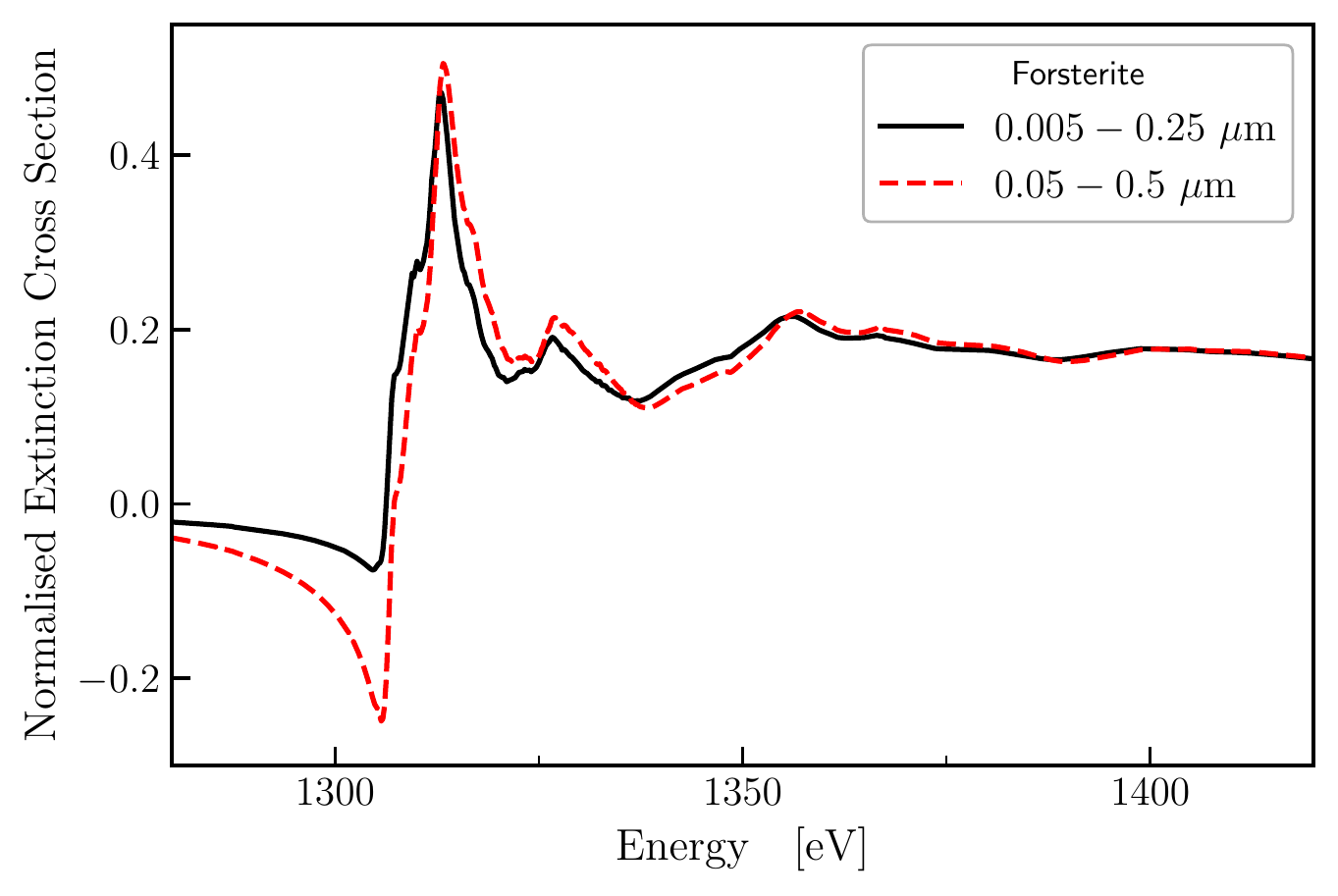}
      \caption{Normalised extinction cross section for forsterite using two different grain size distributions. The solid black line represents the standard MRN grain size ranging between $0.005-0.25\, \mu\rm{m}$. The dashed red line delineates the larger grain size spanning between $0.05-0.5\, \mu\rm{m}$.
              }
         \label{fig:large}
   \end{figure}

\noindent
In Figure \ref{fig:chemistry} we compare, for illustrative purpose, the extinction cross sections of three representative compounds in our sample set: an amorphous Mg-pure pyroxene (blue dashed line), a crystalline Mg-pure olivine (red solid line), and the magnesium aluminium spinel (black solid line). The extinction cross sections of all the compounds are shown in Appendix \ref{app:all_edges}.\\
The chemical properties of the grains, in particular the length of the atom boundaries, determine the shape of the extinction profiles. Both olivine and pyroxene minerals, which share the same silicate tetrahedron, show similar patterns. Spinel, which instead presents an aluminium cubic structure, shows a distinct extinction shape and the main peak of the cross section is shifted at an higher energy. The extinction cross section is also sensitive to the crystalline order of the mineral. Pure crystalline compounds shows multiple, distinct, and narrow peaks, whereas the extinction profiles of amorphous compounds is smoother and does not show any secondary peaks. 

\subsection{Large grain size distribution}
\label{sec:large}
In order to investigate the grain size, in particular focusing on the presence of particles larger than $0.25\ \mu\rm{m}$, we calculate and implement in \texttt{amol} the extinction cross sections adopting a modified MRN grain size distribution. Specifically, we adopt the distribution already presented by \cite{Zeegers17} with $(0.05 \le a \le 0.5)\ \mu\rm{m}$. In Figure \ref{fig:large} we show the effect of this change in the grain size distribution. The forsterite Mg K-edge with a MRN distribution with particle size of $0.005 - 0.25\ \mu\rm{m}$ is shown in black and in red the same edge but now with a $L$MRN size distribution that has a particle size range of $0.05-0.5\ \mu\rm{m}$. The feature before the edge ($\sim 1305\, \rm{eV}$), namely the scattering peak, is sensitive to the dust grain size and it is due to an enhanced scattering efficiency for larger grain size \citep[][]{Zeegers17,Rogantini18}.

\subsection{Magnesium and silicon models}
In this paper, for the first time, we simultaneously analyse the magnesium and the silicon K-edges of a bright LMXB. We built our extinction model joining the laboratory cross sections of the Mg K-edge, from the present work, together with the Si K-edge cross sections taken from \cite{Zeegers17,Zeegers19}. The extinction models with both Mg and Si K-edges have been implemented in the \texttt{amol} model. Moreover, we included in our models the Si-bearing compounds that do not contain magnesium in the molecules, such as quartz and fayalite. We also add magnesium oxide (also known as magnesia) which contain only magnesium \citep{Fukushi17}. Since the Si K-edge model of hypersthene is not available \citep{Zeegers19}, we do not use this compound during the analysis and we just present its Mg K-edge cross-section. Moreover, we do not include spinel in the fitting of the astronomical data. This is because the aluminium in the compound could be misquantified due to possible calibration residuals in the Al edge\footnote{see Figure 1.3 of \textit{"The Chandra Proposers’ Observatory Guide"} version 20.0 (\url{http://cxc.harvard.edu/proposer/POG/}).} of the source spectrum.\\
We report in Table \ref{tab:sample} (with the index \#Mod) the complete list of the extinction profiles used to analyse the Mg and Si K-edges interstellar medium absorption observations.


\section{Astronomical observation}
\label{sec:esa}

\subsection{GX 3+1}

We use as a test source the bright X-ray binary GX 3+1. It has a persistent bolometric luminosity of $\sim 6 \times 10^{37}\, \rm{erg}\, \rm{s}^{-1}$ \citep{denHartog03}, and the spectrum shows deep Mg and Si absorption edges at 1.308 and 1.840 keV (9.50 and 6.74 {\AA}), respectively. GX 3+1 (also known as Sgr X-1 and 4U 1744-26) is one of the first discovered cosmic X-ray sources. It was detected during an \emph{Aerobee}-rocket flight on June 16, 1964 \citep{Bowyer65}. Ever since, it has already been intensely observed with multiple satellites: HAKUCHO \citep{Makishima83}, GRANAT \citep{Lutovinov03}, GINGA \citep{Asai93}, RXTE \citep{Kuulkers00}, \textit{Beppo}-SAX \citep{Oosterbroek01,denHartog03,Seifina12}, INTEGRAL \citep{Chenevez06}, XMM-\textit{Newton} \citep{Piraino12,Pintore15} and \emph{Chandra} \citep{Schulz16}. The detection of multiple thermonuclear bursts \citep{Makishima83,Kuulkers02a} suggests that the compact object hosted in GX 3+1 was an accreting neutron star. Thanks to the detection of these X-ray bursts with radius expansion the distance to the source was estimated to be in the range $4.2-6.4\ $kpc with a best estimate of $\sim 6.1\, \rm{kpc}$ \citep{Kuulkers00,denHartog03}. Spectral analysis of the source showed that its X-ray spectrum can be described by a model comprised of a blackbody component, most likely associated with the accretion disc, and a Comptonized component, produced by an optically thick electron population located close to the neutron star corona \citep{Oosterbroek01,Mainardi10,Seifina12}.

\subsection{Data reduction}

We use here seven datasets of \emph{Chandra} (see Table \ref{tab:continuum}), taken in timed exposure (TE) mode between July 2014 and May 2017, for a total exposure of $\sim 213\,\rm{ks}$. The spectrum has been observed by the ACIS-HETG instrument of \emph{Chandra} \citep{Canizares05}. Each dataset contains both HEG and MEG grating spectra which have been downloaded from the \emph{Chandra} Grating-Data Archive and Catalogue \citep[TGCat,][]{Huenemoerder11}. Using the \emph{Chandra} Interactive Analysis of Observations \citep[CIAO,][]{Fruscione06} we combined the +/- first order for each HEG and MEG observation. HEG and MEG spectra of a single observation are fitted together with the same model, correcting when necessary, their instrumental normalisations. In total, we fit simultaneously 14 spectra.\\
The average count rate is $\sim 100 $ counts per second, which translates to a flux of $\sim 4.7 \times 10^{-9}\ \rm{erg}\ \rm{s}^{-1}\ \rm{cm}^{-2}$ in the range $2-10\ $keV \citep{Oosterbroek01}. Because of this high flux the observations are affected by pile-up. The bulk of the pileup photons comes from the MEG first order where the Si K edge resides on a back illuminated CCD. HETGS has a high effective area between 1 and 3~keV, and we exclude some of these data ($E>1.55\,\rm{keV}$). For HEG we consider the broad energy band in the range $1.1-5.2\,\rm{keV}$ ($\sim 2.4-10.8\, $~\AA, respectively).

\subsection{Continuum}
\label{sec:continuum}

In order to represent the continuum of GX 3+1 we assume the presence of both thermal and non-thermal emission \citep{Mitsuda84}. Among the thermal components present in \textsc{Spex} we test a black body \citep[\texttt{bb},][]{Kirchhoff60}, a disk-black body \citep[\texttt{dbb},][]{Shakura-Sunyaev73,Shakura73} and a black body modified by Compton emission \citep[\texttt{mbb}][]{Rybicki86,Kaastra89}. The tested non-thermal components are power law (\texttt{pow}) and Comptonization model \citep[\texttt{comt},][]{Titarchuk94}. The best fit model for GX 3+1 shows a black body plus a power-law, absorbed by a cold absorbing neutral gas model, simulated by the \texttt{hot} model in \textsc{Spex} \citep{dePlaa04,Steenbrugge05}. \\

\noindent
For the cold absorption model, we fix the electron temperature at the lower limit of the \texttt{hot} model, that is $T_e = 0.5\, \rm{eV}$. We update the photo-absorption cross section of neutral magnesium in \textsc{Spex}, adding the resonance transitions, $ 1s\rightarrow np$, calculated using the Flexible Atomic Code\footnote{Flexible Atomic Core, or FAC, is a software package to calculate various various atomic radiative and collisional processes, including photo-ionization and auto-ionization \citep{Gu08}.} (A. Raassen, private communication). Our neutral Mg K-shell cross section is consistent with the Mg \textsc{i} profile obtained by \cite{Hasoglu14} applying the $R-$matrix method.\\

\begin{figure}
   \centering
   \includegraphics[width=\hsize]{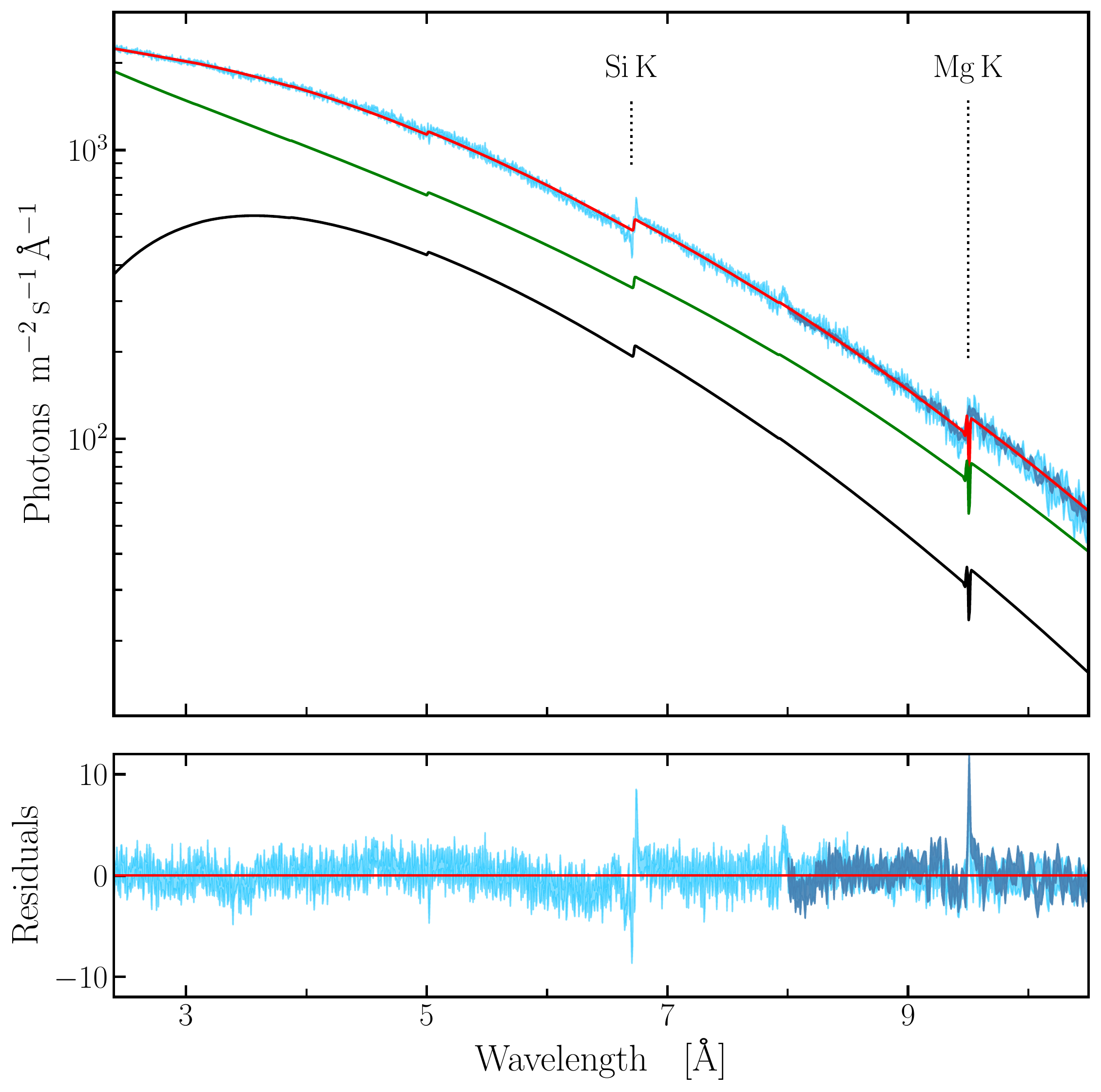}
      \caption{The continuum of GX 3+1. HEG (light blue) and MEG (dark blue) data from seven datasets were used to fit the continuum. We stacked and binned the observations for display purpose only. The average fit of the seven data sets is shown with a red solid line. The model consists of a black body (in black) and a power-law (green) with a photo-electric absorption component with $N_{\rm{H}} \simeq 1.9\times 10^{22}\,\rm{cm}^{-2}$. In the bottom panel we show the relative residuals defined as $(observed-model)/error$. The residual around 8 {\AA} is due to the uncertainty on calibration for the aluminium K-edge.
              }
         \label{fig:contiunuum}
   \end{figure}

\noindent
We fit simultaneously the multiple datasets by coupling the absorption by neutral gas in the interstellar matter that we assume constant. The model is fitted to the data using the $C$-statistic \citep{Cash79}. Using the abundances tabulated by \cite{Lodders10}, we obtain a hydrogen column density $N_{\rm{H}} = (1.91 \pm 0.05)\times 10^{22}\, \rm{cm}^{-2}$ consistent with the values of previous works. The average best fit of the continuum is represented in Figure \ref{fig:contiunuum} and the parameter values for each observation are reported in Table \ref{tab:continuum}. The residuals in the Mg and Si K-edge region hint that we are overestimating the content of these two elements in gas phase. Thus, it is necessary to add the dust model in order to fit the residuals present. \\ 
Furthermore, we test the presence of collisionally ionised gas along the line of sight and gas outflow from the source in its environments. Thus, we add to our model an extra \texttt{hot} component plus the photo-ionised absorption model \citep[\texttt{xabs} in \textsc{Spex}][]{Steenbrugge03}. We do not find any evidence of ionised gas along the line of sight.

\begin{table*}[t]
\centering 
\small{
\caption{Broad band modelling of GX 3+1 using HEG and MEG data of seven observations.}             
\label{tab:continuum} 
\renewcommand{\arraystretch}{1.2}     
\begin{tabular}{c | c c c | c c c c c | c | c }     
\hline\hline       

\multirow{2}{*}{\#} & \multirow{2}{*}{Obsid} & Date & $t_{\rm{exp}}$ & $N_{\rm{H}}$ & N$_{\tt{pow}}$ & $\gamma$ & N$_{\tt{bb}}$ & $T_{\tt{bb}}$ & $F_{2-10\, \rm{keV}}$ & \multirow{2}{*}{Cstat/d.o.f}\\
& & \scriptsize{mm/yyyy} & \scriptsize{ks} & \scriptsize{$10^{22}\ \rm{cm}^{-2}$} & \scriptsize{$10^{44}\ \rm{ph}\,\rm{s}^{-1}\,\rm{keV}^{-1}$} & \scriptsize{-} & \scriptsize{$10^{13}\ \rm{cm}^{2}$} & \scriptsize{keV} & \scriptsize{$10^{-9}\,\rm{erg}\,\rm{cm}^{-2}\,\rm{s}^{-1}$}\\ 
\hline                    
   1 & 16307 & 07/2014 & 43.59 & \multirow{7}{*}{$1.91\pm0.02$} & $29\pm2$ & $1.12\pm0.04$ & $2.7\pm0.1$ & $0.8\pm0.1$ & $7.5\pm0.5$ & $4198/3885$ \\  
   2 & 16492 & 08/2014 & 43.59 &                                & $28\pm2$ & $1.09\pm0.04$ & $2.7\pm0.2$ & $0.8\pm0.1$ & $7.4\pm0.5$ & $4206/3885$ \\
   3 & 18615 & 10/2016 & 12.16 &                                & $25\pm2$ & $1.25\pm0.04$ & $1.8\pm0.2$ & $0.7\pm0.2$ & $4.8\pm0.3$ & $4132/3885$ \\
   4 & 19890 & 05/2017 & 29.08 &                                & $29\pm2$ & $1.23\pm0.03$ & $2.1\pm0.1$ & $0.8\pm0.2$ & $6.1\pm0.4$ & $4089/3885$ \\
   5 & 19907 & 11/2016 & 26.01 &                                & $22\pm1$ & $1.21\pm0.04$ & $2.4\pm0.2$ & $0.7\pm0.1$ & $4.8\pm0.3$ & $4058/3985$ \\
   6 & 19957 & 04/2017 & 29.08 &                                & $31\pm2$ & $1.20\pm0.03$ & $2.0\pm0.1$ & $0.8\pm0.2$ & $6.9\pm0.4$ & $4115/3885$ \\
   7 & 19958 & 05/2017 & 29.08 &                                & $29\pm2$ & $1.22\pm0.03$ & $2.1\pm0.1$ & $0.8\pm0.1$ & $6.2\pm0.4$ & $4104/3885$ \\
\hline
\multicolumn{4}{c}{Average} & $1.91 \pm 0.02$ & $28\pm2$ & $1.19\pm0.04$ & $2.3\pm0.1$ & $0.8\pm0.1$ & $6.2\pm0.4$ & $28906/27201$ \\
\hline         
\end{tabular}
}
\tablefoot{A part of the $N_{\rm{H}}$, the parameters are expressed using the default unit of \textsc{Spex}. Errors given on parameters are $1\sigma$ errors.}
\end{table*}

\subsection{Fit of the magnesium and silicon edges}

\begin{figure*}
  \centering
  \includegraphics[width=1.05\hsize]{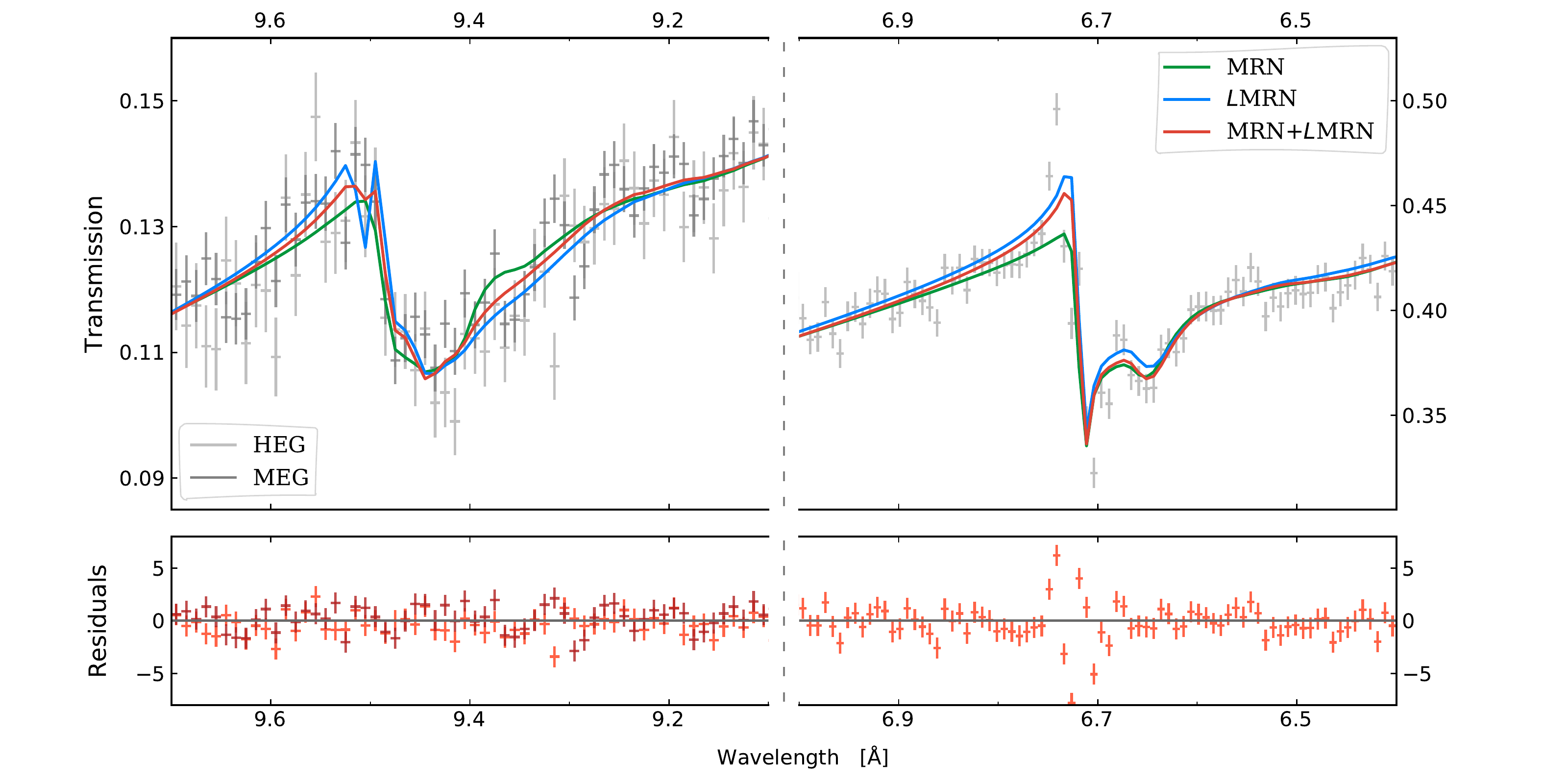} 
  \caption{\textit{Top panel}: the magnesium and silicon K-edges of GX 3+1. The HEG and MEG data are respectively shown in light and dark grey. We do not consider MEG data for the Si K-edge because of the pile-up contamination. We fit the two edges using models with different grain size distributions: MRN models only in green, $L$MRN models only in blue and both MRN and $L$MRN models in red. \textit{Bottom panel}: we show the residuals defined as $(observed-model)/error$ of the best fit obtained using MRN and $L$MRN models. The HEG and MEG data are respectively shown in light and dark red. The data are stacked and binned for display purpose.
          }
  \label{fig:cstat}
   \end{figure*}

After studying the continuum we fit our dust models to \emph{Chandra}-HETG data in order to study the solid phase of the interstellar medium along the line of sight.
In the fit we keep the temperature and the power-law index of the continuum model, leaving the respective normalizations free to vary.\\
The dust models necessary to characterise the near-edge features of the Si and Mg K-edges are implemented in the multiplicative component \texttt{amol}. This \textsc{Spex} model can fit a dust mixture consisting of four different types of dust at the same time. We follow the same method as described in \cite{Costantini12}, where they test all the possible configurations of the dust species and compare all the outcomes using a criterion based on the $C$-statistics value (see Section \ref{sec:model_comp}). The number of compound combinations is given by 
\begin{equation}
\label{eq:combination}
C_{n,k} = \frac{n!}{k!(n-k)!}\quad ,
\end{equation}
where $n$ is the number of compounds, and $k$ the combination class. Since each combination describes a single extinction model, $C_{n,k}$ represent the number of models utilized to fit the data. \\ 
Initially, only the extinction models obtained assuming the MRN size distribution ($n=15$) were selected to study the XANES profiles at the Mg and Si K-edges. In Figure \ref{fig:cstat} we show the best fit of the two X-ray edges with the green solid line. The $C$-statistic ($C$stat) value is 28199 with 27207 degrees of freedom ($d.o.f.$). The mixture that fits the data best consists mainly of amorphous olivine ($\sim 85\% $) and a smaller contribution of magnesium oxide.\\
Furthermore, we test a different size distribution fitting the data using exclusively extinction cross sections obtained adopting the $L$MRN size distribution ($n=15$), presented in Section \ref{sec:ext}. The relative best fit, with a $C$stat/dof equivalent to 28182/27207, is shown in Figure \ref{fig:cstat} with the blue solid line. 
The C-stat improves with LMRN. However this fit requires a large amount of gas, for both Mg and Si, in order to fit the data (30\% and 40\%, respectively). This is difficult to reconcile with literature values \citep{Jenkins09}. \\
In the final analysis we consider both dust size distributions, for all our measurements ($n=30$). The best fit, with a $C$stat value of 28129/27207, is represented with the red solid line. The relative residuals (for both HEG and MEG) are shown in the bottom panel. The dust that best represents the data is a mixture of amorphous olivine ($\sim 71\%$), crystalline fayalite ($\sim 16\%$) and amorphous quartz ($\sim 13\%$). The contribution of MRN and $L$MRN size distribution models is comparable, $\sim 57\%$ and $\sim 43\%$, respectively.\\ 
The Si K-edge shows further residuals around the energy threshold. We further discuss these residuals in Section \ref{sec:sik}.\\
The mixture of standard and large MRN grains gives the best representation of the Mg and Si K-edges. The parameter values for the LMRN+MRN, MRN and LMRN model with their statistical errors are summarised in Table \ref{tab:fit}. For clarity we divide the table in blocks. In the upper block of the table $N_{1-4}$ indicate the column density of each dust species present in the model in units of $ 10^{17}$ cm$^{-2}$. In the second block we list the gaseous phase column density of each element of interest ($N_{\rm{X}}$).\\
We summarise, in Table \ref{table:abu_dep}, the depletion values and total abundances for oxygen, magnesium, silicon and iron. The abundances are calculated considering the total amount of atoms in both gas and solid phase and they are compared with the solar abundances from \cite{Lodders10}.

\subsection{Evaluating the goodness of fit}
\label{sec:model_comp}

Considering all the models calculated using both MRN and $L$MRN size distributions, we obtain 27405 models (from Equation \ref{eq:combination}). The $C$-statistics values representative of different dust mixtures can be similar. Since our candidate models are not-nested and with same number of free parameters, the standard model comparison tests (e.g. the $\chi^2$ Goodness-of-Fit test, the Maximum Likelihood Ratio test, and the F-test) cannot be used to evaluate the significance of the models \citep{Protassov02}. The \emph{Aikake Information Criterion} (AIC)\footnote{AIC is founded in information theory. We refer to \citep{Liddle07} and \citep{Ranalli17} for a extensive introduction to the informative criteria from an astrophysics viewpoint.} represents an elegant estimator of the relative quality of not-nested models without relying on time-consuming Monte Carlo simulations \citep{Akaike74,Akaike98}. The $AIC$ value of a model is defined as
\begin{equation}
\label{eq:aic}
AIC = 2k - 2\ln(\mathcal{L}_{\rm{max}})\,,
\end{equation}
where $k$ is the number of fitted parameters of the model and $\mathcal{L}_{max}$ maximum likelihood value. Recalling the relation $Cstat=-2 \ln\mathcal{L}$ \citep{Cash79} the relation between $C$-statistic and $AIC$ is clear.\\
In our work it is not the absolute size of the AIC value, but rather the difference between AIC values ($\Delta AIC$), that is important. The AIC difference, defined as 
\[
\Delta AIC_{i} =  AIC_{i} - AIC_{\rm{min}}\,,
\]
allows both a comparison and a ranking of the candidates models. For the model estimated to be best, $\Delta AIC_{i} \equiv \Delta AIC_{\rm{min}} \equiv 0$. Following the criteria presented in \cite{Burnham02}, we consider competitive with the selected best model the models with $\Delta AIC_{i} < 10$.\\

\begin{figure}
  \centering
  \includegraphics[width=\hsize]{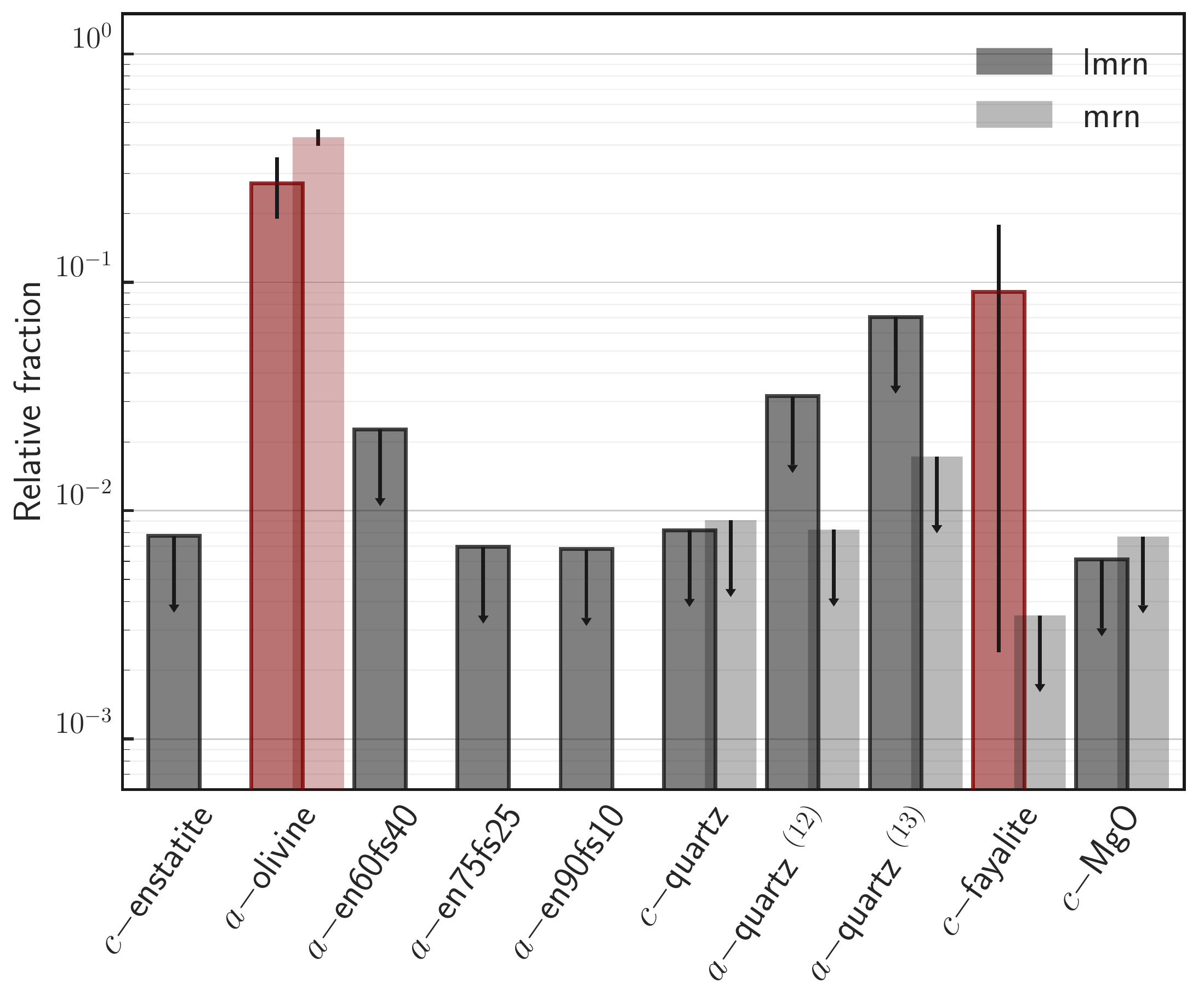} 
  \caption{Bar plot of the relative abundance for each dust species calculated considering $AIC-$selected models. Darker bars represent models with a $L$MRN size distribution instead lighter bars refer to MRN models. With red filled bars, we highlight the dust species with a constrained relative fraction.
          }
  \label{fig:columns}
   \end{figure}

\noindent
From the $AIC$-selected models, we obtain the relative contribution of each dust compound over the total dust obscuration. In Figure \ref{fig:columns} we show the relative fraction of the dust species for both MRN (lighter-colour) and $L$MRN (darker colour) dust size distributions. The red-highlighted bar indicates the compounds for which we are able to constrain their relative contribution. The amorphous olivine is the most representative compound among the selected models, with a relative value of $0.70\pm0.09$ ($0.43 \pm 0.04$ and $0.27 \pm 0.08$ for MRN and $L$MRN size distribution, respectively). In particular, the amorphous olivine ($a-$olivine) is the major contributor for every $AIC$-selected model. Models without any important contribution from $a-$olivine show $\Delta AIC_{i} > 35$. \\
A secondary contribution is given by the crystalline fayalite, with a $L$MRN size distribution, which represents a relative value of $0.091 \pm 0.088$. For the remaining compounds we obtain upper limits (grey bars in Figure \ref{fig:columns}) of their contributions, which are always lower than $0.07$. Regarding the compounds listed in Table \ref{tab:sample}, and missing in Figure \ref{fig:columns} (in explicit, $a$-enstatite, $c$-En60Fs40, $c$-En90Fs10, $a$-forsterite and $c$-olivine), they do not occur in any of the selected models and we consider their contributions negligible in this fit.\\
The models selected with the AIC method agree not surprisingly with the best fit obtained with the $C$stat.
\newline

%
\begin{table}
\caption{Dust and gas column densities obtained by fitting the Mg and Si K-edges.}             
\label{tab:fit}      
\centering
\renewcommand{\arraystretch}{1.3}
\small{                          
\begin{tabular}{r | c c c | c}     
\hline\hline                 
 & \textbf{MRN} & \textbf{$L$MRN} & \textbf{MRN+$L$MRN} & Units\\   
\hline                       
  \multirow{2}{*}{$N_1$} & $c$-quartz                   & $a$-enstatite         & $a$-quartz$^{mrn}$              & \multirow{8}{*}{{\tiny $10^{17}\,\rm{cm}^{-2}$}}                           \\    
                         & $<0.4$                       & $1.9\pm0.8$           & $1.1\pm0.2$                     &                           \\
  \multirow{2}{*}{$N_2$} & $a$-quartz                   & $c$-fayalite          & $a$-olivine$^{mrn}$             &                           \\     
                         & $2.7^{+0.6}_{-0.2}$          & $1.7\pm 0.5       $   & $3.6\pm0.3$                     &                           \\
  \multirow{2}{*}{$N_3$} & magnesia                     & $a$-olivine           & $c$-fayalite$^{lmrn}$           &                           \\      
                         & $1.4^{+0.8}_{-0.3}$          & $2.7^{+0.6}_{-0.1}$   & $1.3\pm 0.3$                    &                           \\
  \multirow{2}{*}{$N_4$} & $a$-olivine                  & $c$-en60fs40          & $a$-olivine$^{lmrn}$            &                           \\      
                         & $5.7^{+0.1}_{-0.2}$          & $<0.2$                & $2.3\pm 0.3        $            &                           \\
\hline
$N_{\rm{O}}$   & $0.9\pm0.2       $ & $1.2\pm0.5         $   & $0.8\pm 0.1         $ & {\tiny $10^{19}\,\rm{cm}^{-2}$} \\
$N_{\rm{Mg}}$  & $<0.2            $ & $2.4\pm0.3         $   & $0.7\pm  0.3        $ & {\tiny $10^{17}\,\rm{cm}^{-2}$} \\
$N_{\rm{Si}}$  & $<0.9            $ & $3.7\pm0.3         $   & $0.6\pm  0.5        $ & {\tiny $10^{17}\,\rm{cm}^{-2}$} \\
$N_{\rm{Fe}}$  & $<2.8            $ & $< 2.0             $   & $0.2^{+2.0}_{-0.1}  $ & {\tiny $10^{16}\,\rm{cm}^{-2}$} \\
\noalign{\vskip 0.7mm}
\hline
\noalign{\vskip 0.5mm}
$Cstat$         & 28199 & 28182 & 28129  & \\
$d.o.f$         & 27207 & 27207 & 27207 & \\
\hline                                  
\end{tabular}
\tablefoot{ {\tiny We use the abbreviations $c-$ for crystalline and $a-$ for amorphous.}}
  }
\end{table}

\section{Discussion}
\label{sec:discussion}

\subsection{Silicon edge residuals}
\label{sec:sik}

The HETG/\emph{Chandra} data in GX 3+1 exhibit a peculiar shape of the pre-edge region of silicon K transition. In Figure \ref{fig:sik} we show a zoom-in of the silicon K-edge. The complexity of this edge was already observed by \cite{Schulz16}. With our larger set of observations (approximately a double exposure time) we observe this complex structure with a significance $\gtrsim 5 \sigma$. \cite{Schulz16} infer that the peak centred at 6.740 {\AA} is due to X-ray scattering. However, its wavelength does not correspond to the scattering peak of our dust extinction model which is instead centred at 6.728 \AA. It is possible that the peak is contaminated by an unresolved and unknown instrumental artefact \citep{Miller02}. Regarding the absorption present right before the onset of the edge we speculate that its origin is attribute to interstellar medium present along the line of sight and we test different possibilities.

\subsubsection*{\it Neutral silicon}
\begin{figure}
  \centering
  \includegraphics[width=\hsize]{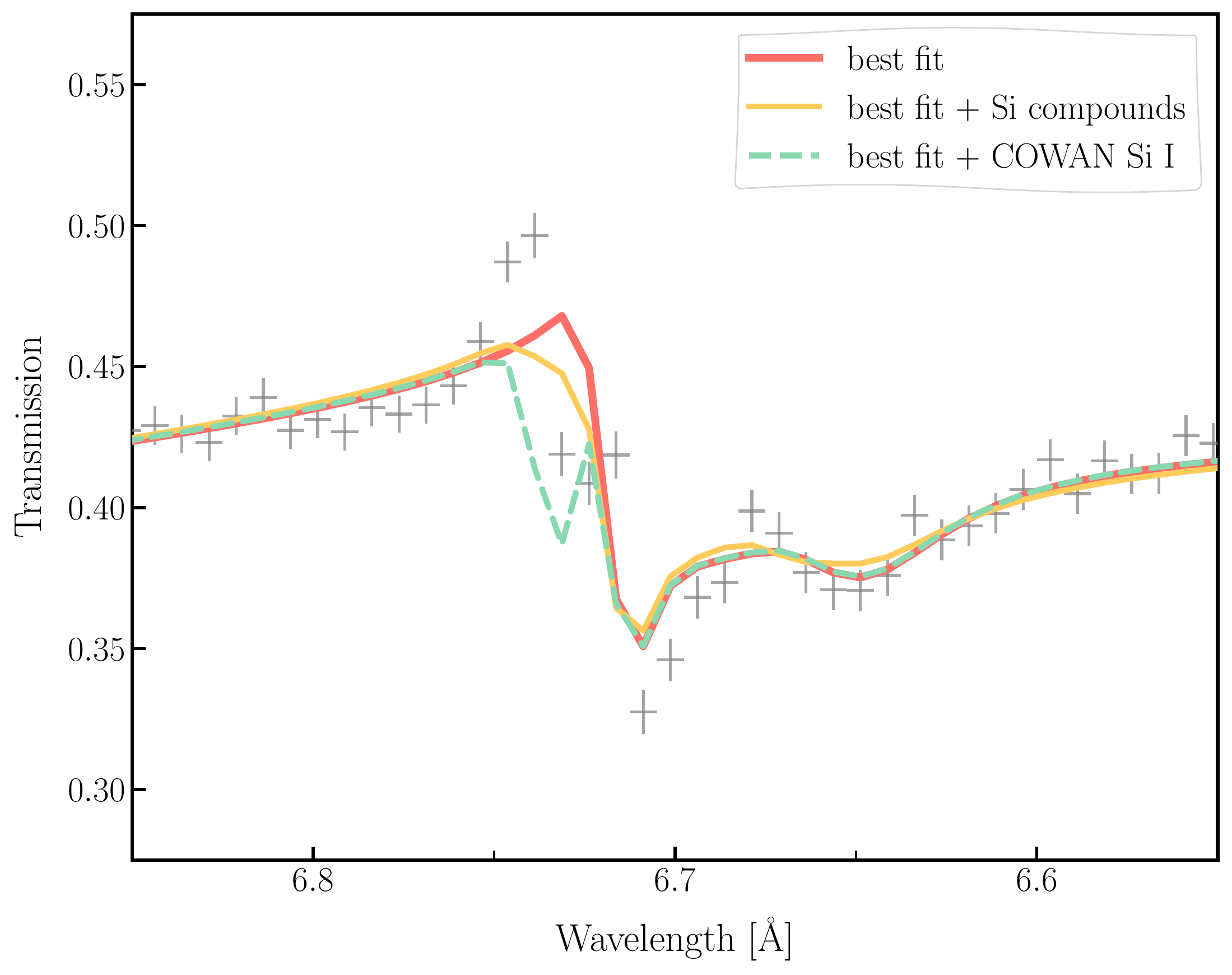} 
  \caption{Zoom-in of the silicon K-edge. We represent the best fit with a red solid line. In orange we show the model obtained by adding secondary Si-bearing dust candidates presented in the text to the best fit. The green dashed line represents the best fit adding the neutral silicon cross section calculated with the COWAN code. We use a data bin size of $\sim 0.04\,$\AA, approximately \nicefrac{1}{3}FWHM (full width half maximum, in other words the energy resolution of the detector).
          }
  \label{fig:sik}
   \end{figure}

The K-shell X-ray absorption for a single, isolated silicon atom presents multiple resonance transitions $1s \rightarrow 3p$. We update the \cite{Verner96} silicon cross section present in \textsc{Spex} with these transitions, calculated using both \textsc{FAC} and \textsc{COWAN} codes \citep[][A. Raassen, private communication]{Cowan81}. Assuming the silicon depletion value found in the best fit ($\delta_{\rm{Si}} = 0.94 $), with the update cross section we obtain an absorption feature with a strength similar to the absorption feature observed right before the onset of the Si K-edge. However, none of these absorption features corresponds exactly to the energy measured by HETG: the absorption line calculated with the two different codes are shifted to lower energies (higher wavelengths) of $\Delta E \sim 1.5-4.5$ eV ($\Delta \lambda \sim 0.006-0.017$ \AA). These shifts are noticeable since the differences are close to the energy resolution of HEG in the silicon region ($\Delta \lambda = 0.012 $ \AA). In Figure \ref{fig:sik} the green dashed line shows the absorption line due to the resonance transitions calculated with the COWAN code, which presents the less divergent shift. Moreover, \cite{Hasoglu18} calculate the K-shell photoabsorption of neutral silicon using a modified version of the $R$-Matrix method \citep{Berrington95}. Their final result is somewhat consistent to our calculation using the COWAN code.

\subsubsection*{\it Ionised gas}

We test if a photoionised gas is able to reproduce the absorption feature right before the onset of the edge. Thus, we add a photoionized component (\texttt{xabs} in \textsc{Spex}) with a systematic velocity that is free to vary to our model. It results in a modest column density ($N_{\rm H} = 3^{+7}_{-2}\times 10^{19}\, \rm cm^{-2}$) for an ionisation parameter of $\log \xi = 1.7^{+0.2}_{-0.4}$. \\
We also test collisional ionized plasma (component \texttt{hot} in \textsc{Spex}), with a temperature constrained between 0.3 and 2 keV, in order to ensure absorption by Si\textsc{xiii}, in the Si K-edge region obtaining an upper limit for the column density ($N_{\rm H}<1.5\times10^{20} \rm \ cm^{-2}$).

\subsubsection*{\it Si-bearing dust}

In the interstellar dust, silicon is potentially able to create a chemical bound with elements different than oxygen, the standard characteristic bond of silica and silicates. Therefore, we included in our Si K-edge model set cosmic dust candidates like crystalline silicon \citep{Witt98,Li02}, crystalline silicon carbide \citep[SiC, ][]{Whittet90,Min07}, and silicon nitride \citep[Si$_3$N$_4$,][]{Jones07}\footnote{The XANES profiles of crystalline Si, SiC and Si$_3$N$_4 $were taken from \cite{Chang99} and analysed with the method present in Section \ref{sec:ext}. We create the extinction model adopting both MRN and $L$MRN size distributions.}. The Si-Si, Si-C and Si-N bonds are characterised by lower-energy thresholds and consequently, their Si K-edges is wavelength-shifted with respect to the silicate's ones. We add magnesia (MgO) to our model in order to compensate for any silicate-poor (and therefore magnesium poor) fitting that we are testing. The resulting model is presented in Figure \ref{fig:sik} with a orange solid line. The fit has a better $C-$statistic value but a lower $AIC$ value due to the penalty term $2k$ in Equation \ref{eq:aic}.

\subsubsection*{\it Double silicate edge}

Adding naively to our best model a supplementary silicate component with systematic velocity free to vary, results in a relative large amount of cold material ($N_{\rm H} \sim 3 \times 10^{17}\, \rm cm^{-2}$) with a receding velocity $\gtrsim 500\, \rm km/s$ along the line of sight. This scenario is hard to justify since such motion of matter is not observed the literature \citep[e.g.][]{vandenBerg14,Pintore15} and furthermore the magnesium K-edge does not show an evident additional, redshifted feature.
\newline
We can also discard porosity and different grain geometries as possible cause of the residuals since \cite{Hoffman16} show that their effects on the silicon edge profile is negligible. 
Presently, we are not able to characterise the features located next to the onset of the Si K-edge between 6.72 and 6.75 \AA. The comparison of the Mg and Si K-edges detected in different lines of sight will be crucial to understand the nature of it in forthcoming works.

\subsection{Depletions \& Abundances}

By fitting the low energy curvature of the X-ray spectrum of GX 3+1 and adopting the protosolar abundances of \cite{Lodders10}, we find a hydrogen column density of $N_{\rm{H}} = (1.91 \pm 0.02) \times 10^{22}\,\rm{cm}^{-2}$. Using the solar abundances of \cite{Anders89}, adopted by previous authors, this value corresponds to $\sim 1.61 \times 10^{22}\,\rm{cm^{-2}}$ and it is consistent with the neutral column density obtained by \cite{Oosterbroek01} where $N_{\rm{H}} = (1.59 \pm 0.12) \times 10^{22}\,\rm{cm}^{-2}$. 
From the residuals of the continuum analysis in Figure \ref{fig:contiunuum}, it is clear that the silicon and magnesium edges can not be represented only by pure gas absorption and it is necessary to introduce the dust component adding \texttt{amol} to the fitting model. The dust models used in the analysis (see Table \ref{tab:sample}) contain oxygen, magnesium, silicon, and iron. The content of these elements in the solid phase is expressed by the depletion values shown in Table \ref{table:abu_dep}.\\
Silicon is highly depleted along the line of sight of the source: more than $90\%$ is found in solid phase. Similarly, a large fraction of magnesium (more than $80\%$) is included in dust grains. The two elements share the similar depletion values in agreement with the fractional depletion observed by \cite{Dwek16} and \cite{Jenkins09}. The depletions of oxygen and iron in dust are derived indirectly from the model since the absorption edges of these elements fall outside the spectral band. Therefore, no strong conclusions can be drawn regarding these two elements. However, oxygen shows a moderate fractional depletion value ($\delta_{\rm{O}}=0.27\pm0.02$). This value would be consistent with the depletion observed for lines of sight with different neutral column densities \citep[e.g.][]{Costantini12,Pinto13}, confirming the weak correlation between the depletion of oxygen and the physical conditions of the environment \citep[e.g. density and temperature,][]{Whittet02}. Instead, iron seems highly depleted consistently with the values shown by \cite{Whittet02} and \cite{Jenkins09}.\\
The total abundances have been evaluated summing the column densities of the gas and the solid phases for each element. Our values do not depart significantly from the solar values (see Table \ref{table:abu_dep}).

%
%
\begin{table}
\caption{Abundances and fractional depletions.}             
\label{table:abu_dep}                                      
\centering                          
\begin{tabular}{c l l}        
\hline\hline
\noalign{\vskip 0.35mm}              
Element &\multicolumn{1}{c}{$\delta_{Z}$} & \multicolumn{1}{c}{$ A_{Z}/A_{\odot} $} \\
\noalign{\vskip 0.5mm}   
\hline
\noalign{\vskip 0.8mm}                      
   Oxygen     & $0.27^{+0.02}_{-0.02}$  &  $1.01^{+0.03}_{-0.02}$ \\
\noalign{\vskip 1.mm} 
   Magnesium  & $0.89^{+0.10}_{-0.10}$  &  $0.88^{+0.07}_{-0.08}$ \\
\noalign{\vskip 1.mm} 
   Silicon    & $0.94^{+0.06}_{-0.08}$  &  $1.21^{+0.08}_{-0.10}$ \\
\noalign{\vskip 1.mm} 
   Iron       & $0.99^{+0.01}_{-0.17}$  &  $1.13^{+0.17}_{-0.09}$ \\
\noalign{\vskip 1.mm} 
\hline                                
\end{tabular}
\tablefoot{We indicate with $\delta_{Z}$ the fractional depletion of the elements. The term $A_{Z}$ represents the total (dust plus gas) element abundance found analysing the spectrum of GX 3+1. The solar abundances used to calculate the ratio $A_{Z}/A_{\odot}$ are taken from \cite{Lodders10}.} 
\end{table}

\subsection{Dust chemistry}

The simultaneous analysis of the silicon and magnesium edges is in principle able to constrain the typical cations-to-silica ratio (Mg+Fe)/Si of interstellar silicate grains. For the best fit of the data we find that $\rm{(Mg + Fe)/Si} \sim 2$, considering only pyroxene and olivine compounds. This gives $\rm{O/Si} = 4 $, implying that silicates along the line of sight present an olivine-type stoichiometry. Indeed, the amorphous and crystalline olivine, together with the fayalite, are characterised by the orthosilicate anion [SiO$^{4-}_{4}$] and represent $\sim 91\%$ of the dust in our fit.\\
Moreover, the best fit is characterised by $\rm{Mg}/ (\rm{Mg}+\rm{Fe}) = 0.41 \pm 0.02$ (and therefore by $\rm{Mg} : \rm{Fe} = 0.69 \pm 0.05$). Fayalite gives the major contribution of iron for the dust component and it is preferred over the magnesium-rich forsterite. We found a larger presence of iron in silicates with respect to the values observed for different lines of sight \citep[e.g.][]{Costantini12}, wavelengths \citep[e.g.][]{Tielens98,Min07,Blommaert14} and/or environments \citep[e.g. the interplanetary medium][]{Altobelli16}, where Mg-rich silicates are detected. However, we cannot constrain the iron depletion value since we cannot characterise directly and simultaneously the iron absorption edges. \\
The analysis of the $AIC$-selected models shows similar ratios (namely, $\rm{(Mg + Fe)/Si}\approx 1.9$ ; $\rm{Mg}/ (\rm{Mg}+\rm{Fe}) \approx 0.45$ ; $\rm{Mg} / \rm{Fe} \sim 0.8$) in agreement with the nominal best-fitting model.

\subsection{Dust crystallinity}
The best fit suggests the presence of a relatively large amount of crystalline grains along the line of sight of GX 3+1. Defining the crystalline-to-amorphous ratio as $\zeta_{1} = $ crystalline dust / (crystalline dust + amorphous dust) \citep{Zeegers19} we find a value of $\zeta_1 = 0.15\pm0.03 $. This is consistent to the range of values found by \cite{Zeegers19}, $\zeta_1 = 0.04-0.12$, using several several low mass X-ray binaries. Comparing our results with the literature, we find higher percentages of crystalline dust with respect to the fractions observed by the infrared spectroscopy. For example, \cite{Li07}, \cite{Kemper04}, analysing the $9.7\, \mu\rm{m}$ and $18\, \mu\rm{m}$ features, set the possible maximum crystalline fraction of the total silicate mass in the interstellar medium to a maximum of $5\%$ and $1.1\%$, respectively. \\
The crystalline ratio that we find may be partially biased by the limitation of our measured set of compounds. Previous works already showed and discussed the presence of this bias \citep{Zeegers17}. Indeed, our laboratory model set does not include the amorphous counterpart for all the compounds (see Table \ref{tab:sample}) and consequently, the estimation of the crystallinity may be overestimated. Indeed, in our case, the crystalline ratio is led by the crystalline fayalite, for which the amorphous counterpart, which might contribute to the total fit, is not available. \\
However, if this observed amount of crystalline dust is real, the differences with the infrared observations might be explained in two possible ways. The first is that cold and dense regions, accessible only by X-rays, host cosmic dust with a different crystalline order with respect to the grains which populate the diffuse medium. This differences could be explained by the theoretical model presented by \cite{Tanaka10} and \cite{Yamamoto10} which predict a low-temperature crystallization of amorphous silicate grains induced by exothermic chemical reactions. An alternative explanation is that we might be detecting poly-mineralic silicates, which are expected to be agglomerated particles, possibly containing both crystalline and glassy constituents. In this case, because X-rays are sensitive to a short-range order, XAFS would show crystalline features, whereas there might not be sharp crystalline features in the infrared spectrum \cite{Speck11}.

\subsection{Dust size}

The best fit of the magnesium and silicon K edges is obtained assuming the existence of two dust populations with different size distributions: MRN and $L$MRN presented in Section \ref{sec:ext}. From the results of analysing the two X-ray edges, the two distributions show similar weights with $\rm{MRN}/( \emph{L} \rm{MRN} + \rm{MRN})\sim 0.57$.\\
This approach was motivated by the known complexity of the line of sight towards GX 3+1 which is at a distance of about 6.1 kpc \citep{Kuulkers00} and at longitude $l=2.29$ and latitude $b=0.79$ in Galactic coordinates \citep{Ebisawa03}. Therefore, assuming a distance towards the Galactic centre of 8.5 kpc, the source is located in the outskirts of the Galactic Bulge. In particular, it is situated just behind the "near 3 kiloparsec arm", which is expanding at a speed of $\sim 57\, \rm km/s$ \citep{vanWoerden57,Dame08}, and the Molecular Ring \citep{Clemens88,Jackson06}, respectively located at a distance of 5.5 and 5 kpc from the Sun. The CO emission in these latter two regions is clumped \citep{Bania80}. Especially, the Molecular Ring contains about 70\% of all the molecular gas inside the solar circle \citep{Clemens88}. The ring is thus an enormous reservoir of material in gaseous and solid form. Moreover, GX 3+1 is aligned with the known Bania's Clump 2, a molecular cloud complex near the Galactic Center \citep{Bania77,Stark86}. Several dust lanes in the Galactic bar seem to be connected with this feature \citep{Liszt08}. However, because of the uncertainties on the distance estimate, it is unclear if our source is embedded in these structures or if it is in front of them \citep{Marshall08}.\\
Within 5 kpc from the Sun the line of sight crosses several spiral arms \citep[located at different solar distances,][]{Benjamin08,Urquhart14}: Sagittarius-Carina ($ \sim 1.5 \, \rm kpc $), Scutum-Centaurus ($ \sim 3.5 \, \rm kpc $) and Norma arm ($ \sim 4.5 \, \rm kpc $).\\

\noindent
In this scenario, diffuse dust with a MRN distribution would be naturally located within the spiral-arms. The large dust population, containing mainly amorphous olivine and fayalite, would instead probably belong to the molecular regions close to the Galactic Centre. Here, shielded from dissociating interstellar radiation, the dust grains may easily grow \citep[e.g. ][]{Chapman09,Hirashita12}. The presence of large grains in these regions is thought to be supported by the 'core shine' effect seen in mid- and near-infrared observations of dark clouds \citep[e.g.][]{Pagani10,Lefevre14}.\\
Moreover, at these particular coordinates, the 3D polarization model, developed by \cite{Martinez-Solaeche18} taking into account observed dust emission and observed intensity and polarisation power spectra, shows an enhanced polarization vector $P$ at distance $\gtrsim 3.8\, \rm kpc$. As found by \cite{Kim95} the astronomical silicates with size $a>0.1\,\mu \rm m$ can reproduce the observed polarization of the starlight.\\
Finally, previous works have already observed the presence of different dust populations distributed along the line of sight. Recently, \cite*{Vasilopoulos16} analysed the X-ray dust-scattered rings from the LMXB V404 Cyg. They found that the dust grains are concentrated in different dust layers each characterised by a different size distribution. However, future works, using different grain size distributions such as \cite{Weingartner01}, \cite{Zubko04} and THEMIS \citep{Jones13,Kohler14} as well as an in-depth understanding of the instrument around crucial edges, will be necessary in order to study the dust distribution in detail.\\


\section{Conclusions}
\label{sec:conclusion}

In this paper we present the first attempt to build an X-ray broadband extinction model with multiple edges. In particular, in this work we focus on the simultaneous modelling of the magnesium and silicon extinction profiles both based on synchrotron measurements. This approach allows to better constrain the cosmic dust properties and to avoid degeneracies that can occur using a single edge fit.\\
We introduce the Mg K-edge extinction cross sections of 12 different dust species focusing on their XANES profiles. We analyse the X-ray spectrum of the bright LMXB GX 3+1, whose hydrogen column density is optimized for the simultaneous detection of the magnesium and silicon edges, and we characterise the gas and dust along the line of sight. Below we summarise the main results.
\begin{itemize}

\item[$\bullet$] The absorption spectrum shows the presence of both gas and dust along the line of sight. We find standard solar abundances of magnesium and silicon ($A_{\rm Mg}/A_{\odot} = 0.88 \pm 0.08$ and $A_{\rm Si}/A_{\odot} = 1.21 \pm 0.10$) and their relative depletion values are in agreement with the results obtained by \cite{Jenkins09} ($\delta_{\rm Mg} = 0.89 \pm 0.10 $ and $\delta_{\rm Si} = 0.94 \pm 0.6$).

\item[$\bullet$] We find that amorphous olivine is the most representative dust species along the line of sight of GX 3+1. The olivine orthosilicates characterised by the anion SiO$_4$ are highly preferred over silica (SiO$_2$) and pyroxene (SiO$_3$). For the best fit we find a cations-to-anion ratio of $\rm{(Mg + Fe)/Si} \sim 2$.

\end{itemize} 

\noindent
To fully characterise the silicon and magnesium K-edges and in view of future X-ray missions with better resolving power (XRISM and Athena) it is necessary to develop accurate and detailed cross sections of neutral gas. Those can indeed explain the residuals that we still find in our analysis in the Si K-edge region and they will help to determine the depletion values of the respective elements. Moreover it is also crucial to understand if the pre-edge of the silicon K-edge is contaminated by possible instrumental artefacts and how these affect the modelling of dust extinction.  

\begin{acknowledgements}
DR, EC, IP and MM are supported by the Netherlands Organisation for Scientific Research (NWO) through \emph{The Innovational Research Incentives Scheme Vidi} grant 639.042.525. The Space Research Organization of the Netherlands is supported financially by NWO. We acknowledge SOLEIL for provision of synchrotron radiation facilities and we thank Delphine Vantelon for assistance in using the beamline LUCIA. This research has made use of data obtained from the \emph{Chandra} Data Archive and the \emph{Chandra} Source Catalog, and software provided by the \emph{Chandra} X-ray Center (CXC) in the application package CIAO. We thank A.J.J. Raassen for providing the photoabsorption cross section of neutral magnesium and silicon. We are grateful to P. Ranalli for useful discussions. We also thank A. Dekker and D. Lena for reading an early draft of the manuscript and for providing valuable comments and suggestions. 
\end{acknowledgements}

%
%

   \bibliographystyle{aa} 
   \bibliography{../bibliography/biblio}
   

\begin{appendix} 

\section{Mg K-edge shift}
\label{app:mgk_shift}

\begin{figure}
  \centering
  \includegraphics[width=\hsize]{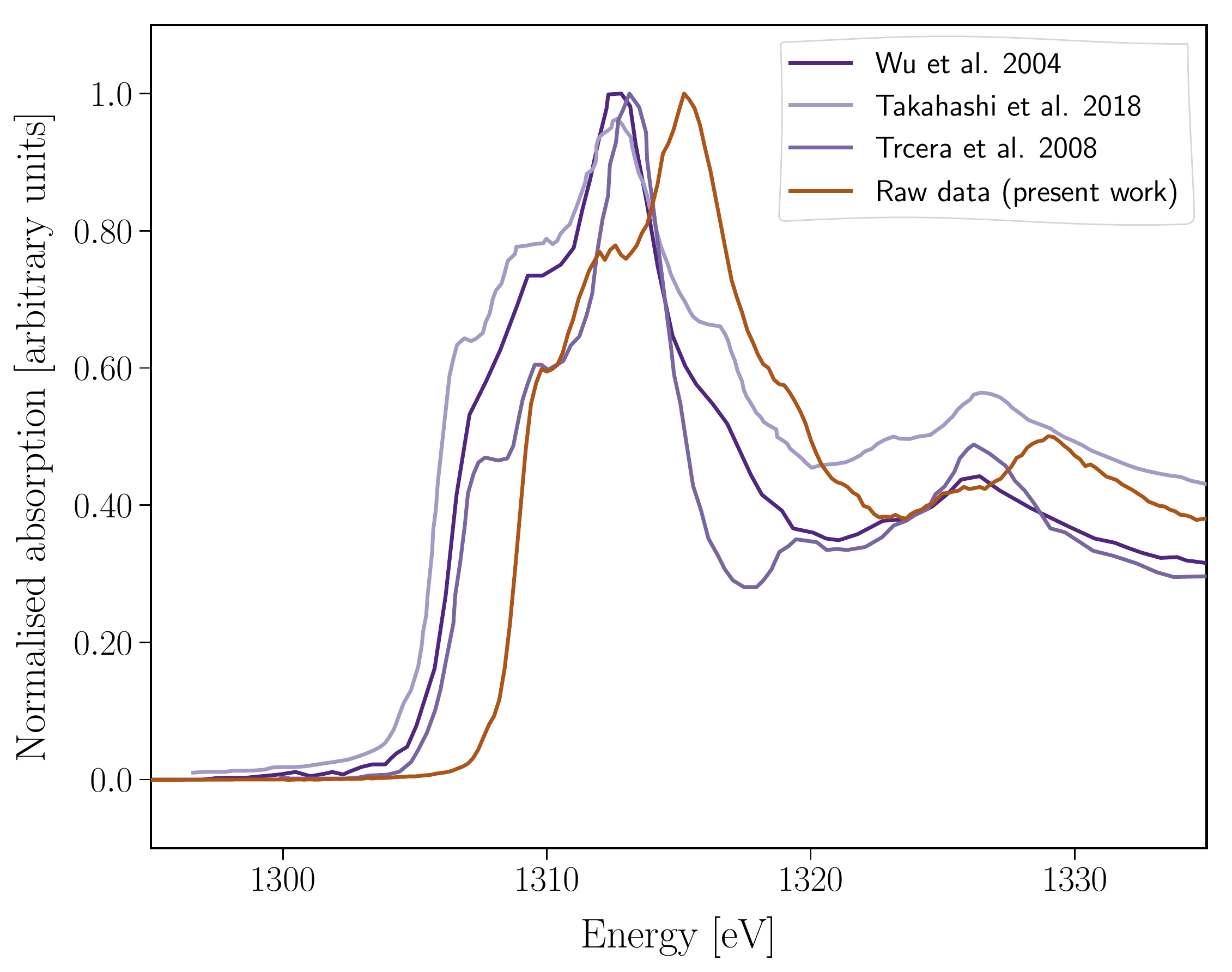} 
  \caption{XANES spectra of crystalline forsterite at the Mg K-edge. Experimental spectra from different works are listed in the panel.
          }
  \label{fig:mg_shift}
   \end{figure}

Comparing our synchrotron measurements with previous and independent works we notice a discrepancy in the energy of the magnesium K-edge threshold. In particular, in Figure \ref{fig:mg_shift} we compare the normalised XANES spectra of crystalline forsterite. We show the result of our measurements with the XANES profiles obtained by \cite{Wu2004}, \cite{Trcera09}, and \cite{Takahashi18}. Our result appears shifted to higher energies with respects to the other reference XANES spectra. We chose to evaluate the energy shift calibrating our model on the spectrum of GX 3+1 presented in this paper in Section \ref{sec:data}. We set the systematic velocity of the absorber (\texttt{zv} parameter in \texttt{AMOL}) as a free parameter and we run the fit of the magnesium K-edge using all the combination of minerals (see Equation \ref{eq:combination}). This approach to define the absolute energy value of a particular transition (difficult to define by experimental data and/or calculations) was already adopted by \cite{Gorczyca13} for the atomic oxygen. We selected the models with a $\Delta AIC<2$ \citep[][]{Burnham02} from the best fit and we found an average speed of $zv=-585\, \rm km/s$ corresponding to an energy shift of $E=-2.54$~eV in the Mg K-edge region. This value is in agreement with previous works shown in Figure \ref{fig:mg_shift}.\\

\section{Extinction cross sections of the Mg K edges.}
\label{app:all_edges}
   \begin{figure*}
            \includegraphics[width=\hsize]{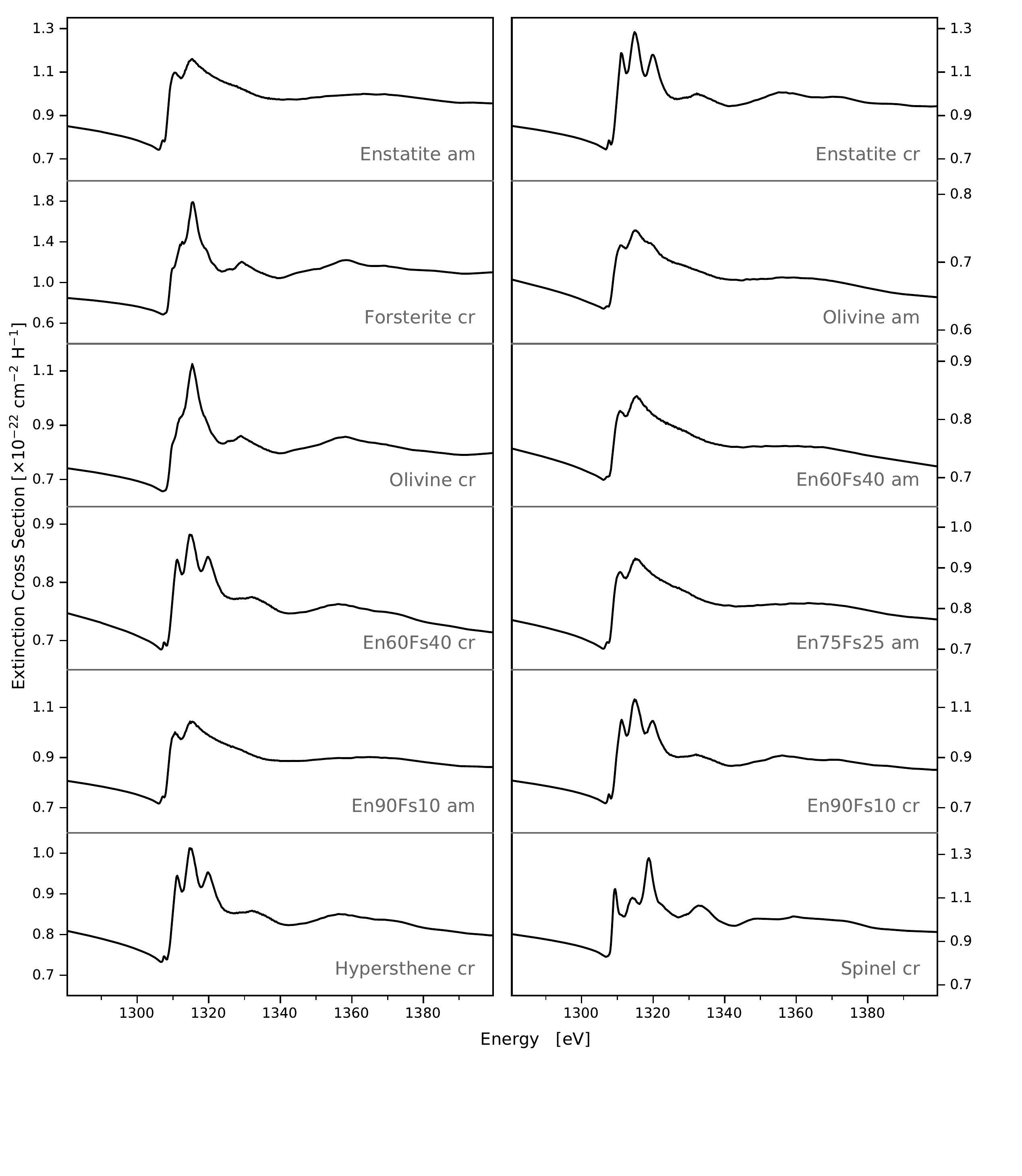}
      \caption{Mg K-edge extinction cross sections of the mineral compounds presented in this work and listed in Table \ref{tab:sample}.
              }
         \label{fig:all_edges}
   \end{figure*}

In Figure \ref{fig:all_edges} we present the extinction cross section profiles around the magnesium K-edge for each sample presented in \ref{tab:sample}. The measurements were taken at SOLEIL (Paris). We adopt the standard MRN size distribution \citep{Mathis77} to obtain the extinction cross sections. These curves were implemented in the \texttt{amol} model of the spectral fitting code SPEX with a fixed energy resolution of 0.1 eV. The absorption, scattering, and extinction cross sections of the compounds (with an energy range between 1100 and 1550 eV) are available in ASCII format at the following links: \url{www.sron.nl/~elisa/VIDI/} and \url{https://zenodo.org/deposit/2790329}.

\end{appendix}

\end{document}